\documentclass[%
 reprints,
 twocolumn
]{revtex4-2}
\usepackage{iceberg}

\begin{document}

\title{Explicit construction of low-overhead gadgets for gates on quantum LDPC codes}
\author{Paul Webster}
\affiliation{Iceberg Quantum, Sydney}

\author{Samuel C.\ Smith}
\affiliation{Iceberg Quantum, Sydney}

\author{Lawrence Z.\ Cohen}
\affiliation{Iceberg Quantum, Sydney}

\author{{\scriptsize\texttt{\{paul,sam,larry\}@iceberg-quantum.com}}}

\begin{abstract}
Quantum low-density parity check (QLDPC) codes can significantly reduce the overhead of quantum computing, provided the methods for performing logical operations do not require substantial space and time resources.
A popular method for performing logical operations is by measuring logical Pauli operators.
We present a simple, explicit construction for fixed gadgets that can measure arbitrary logical Pauli operators on QLDPC codes when dynamically connected to the code block.
We apply this construction to a family of generalised bicycle codes with distances relevant to utility-scale quantum computation ($10\leq d \leq 24$) and show that it reduces the space overhead by at least an order of magnitude compared to corresponding surface code architectures, without increasing the time overhead.
\end{abstract}

\maketitle

\section{Introduction}
Conventional fault-tolerant quantum computing architectures, based on the surface code, have a significant overhead, as they use hundreds or thousands of physical qubits to encode one logical qubit~\cite{Litinski2019}.
This overhead is delaying the realisation of practical applications, such as quantum chemistry or cryptography, since such applications require a quantum computer with a million or more qubits when using such architectures~\cite{Beverland2022,Litinski2023,Gidney2025}.
QLDPC codes, which retain the low check weight of surface codes but allow for non-local connections, could substantially reduce this overhead by encoding many logical qubits in each code block~\cite{Breuckmann2021}, accelerating the timescale for useful applications.

Implementing a universal gate set on QLDPC codes is more challenging than for surface codes, since it requires a method for addressing particular logical qubits within a code block.
This challenge was overcome by a generalisation of lattice surgery that uses ancillary gadgets to measure logical Pauli operators via low-weight check measurements~\cite{Cohen2022}.
When supplemented with magic states, this suffices for universal quantum computation \cite{Litinski2019}.
A number of subsequent works have built on this approach, with a particular focus on reducing its overhead~\cite{Bravyi2024,Cowtan2024-1,Cowtan2024-2,Cross2024,Zhang2025,Williamson2024,Ide2025,Swaroop2025,He2025,Cowtan2025}.
However, it remains an open question how best to implement the approach in practice.

One approach could be to design a bespoke gadget for each logical operator, and to implement this through a dynamic ancilla system that rearranges into a different gadget for each logical cycle.
However, such significant reconfigurations of real hardware may not be practical, especially since they would be conditional on preceding measurement outcomes and so could not be pre-programmed. 

It is therefore preferable to instead design a static gadget capable of measuring arbitrary logical operators.
The extractor system of Ref.~\cite{He2025} is a general framework for this, but it leaves open the question of how to optimally construct an actual low-overhead architecture.
Meanwhile, gadgets have been developed for the gross (and two-gross) code~\cite{Bravyi2024,Cross2024,Williamson2024,Yoder2025}, but these can only measure a subset of logical operators, which leads to a significant time overhead from compilation~\cite{Yoder2025}.
The general problem of how best to design a gadget capable of efficiently measuring all logical Pauli operators on a given QLDPC code therefore remains open.

We address this problem by presenting an explicit construction for low-overhead gadgets with wide applicability.
This involves constructing fixed gadgets for particular seed operators, such that relocating and bridging these gadgets allows for the measurement of arbitrary logical Pauli operators.
For $\llbracket n,k,d \rrbracket$ code families requiring only a constant number of seed operators with weight $O(d)$, the total size of the gadget is then $\tilde{O}(d)$, compared with $\tilde{O}(n)$ for a general extractor system~\cite{He2025}.
By applying our construction to a family of generalised bicycle codes with this property, we show that our construction reduces the overhead by a factor of between $11$ and $24$ compared to surface code architectures of the same distance for the range of distances, $10\leq d \leq 24$, most relevant to early utility-scale quantum computation.

\section{Construction}
In this section, we present our construction for a low-overhead gadget that allows arbitrary logical Pauli measurements on a QLDPC code.
In Sec.~\ref{sec:2.1}, we explain how to construct a gadget for measuring a specific logical Pauli operator.
In Sec.~\ref{sec:2.2}, we use this as the basis for our construction of a fixed gadget that can measure arbitrary logical Pauli operators.

\subsection{Gadget for Measuring a Particular Logical Operator}
\label{sec:2.1}

We explicitly describe a gadget for measuring a logical operator.
For concreteness, we assume that the operator acts as $X$ on all physical qubits in its support.
For an arbitrary logical operator, the gadget would be transformed by the action of single-qubit Clifford gates on the relevant code block qubits.

Consider an arbitrary $X$-type logical operator $L=\prod_i X_{q_i}$ (such that $\{q_i\}=\text{supp}(L)$ is the support of $L$).
Let $\mathcal{S}_L=\{S_j\}$ be the subset of $Z$-type checks of the code that act non-trivially on $\text{supp}(L)$.
A gadget for measuring $L$ can be constructed as follows:
\begin{enumerate}
    \item For each check $S_j\in \mathcal{S}_L$, create a corresponding qubit, $\kappa_j$, in the gadget and add $\kappa_j$ to the support of $S_j$.
    \item For each qubit $q_i\in \text{supp}(L)$, create a corresponding $X$-type check, $\chi_i$, in the gadget. Add $q_i$ to the support of $\chi_i$. For each $j$, add $\kappa_j$ to the support of $\chi_i$ iff $q_i$ is in the support of $S_j$. 
    \item Add $|\mathcal{S}_L|-\text{wt}(L)+1$ $Z$-type gadget checks with the support of each corresponding to a different element of a basis of ker$\left(H_{X,\text{gadget}}\right)$, where $H_{X,\text{gadget}}$ is the parity check matrix for the $X$-type gadget checks, $\{\chi_i\}$ restricted to the gadget qubits $\{\kappa_j\}$.
\end{enumerate}
Steps 1 and 2 construct a gadget whose Tanner graph is the dual of the restriction of the code's Tanner graph to the support of $L$ and connecting each gadget node to its corresponding code node.
Step 3 fixes the additional gauge degrees of freedom that result from adding the gadget.

The measurement outcome of $L_X$ is determined by the product of the measurement outcomes of the gadget checks, $\chi_i$.
Indeed:
\begin{align}
\prod_i\chi_i &=\prod_i\left(X_{q_i}\prod_j \left\{X_{\kappa_j}\mid q_i\in S_j\right\}\right)\\
&=\left(\prod_i X_{q_i}\right) \left(\prod_j\prod_{i} \left\{X_{\kappa_j}\mid q_i\in S_j\right\}\right)\\
&=L\prod_j X_j^{|S_j\cap L|}\\
&=L
\end{align}
where we use that $|S_j\cap L|\equiv 0$ mod 2 (and hence $X_j^{|S_j\cap L|}=I$) for all $j$, since $S_j$ is a $Z$ check of the code which must commute with the $X$-type logical operator $L$.

We can verify that all checks commute, ensuring that the combined code-gadget system is a valid stabiliser code.
Indeed, by construction, $\text{supp}(S_i)\cap \text{supp}(\chi_i)=\{q_i,\kappa_i\}$ and $\text{supp}(S_j)\cap \text{supp}(\chi_i)=\emptyset$ if $i\neq j$, so
$|\text{supp}(S_j)\cap \text{supp}(\chi_i)|=2\delta_{i,j}\equiv 0 \text{ mod 2}$ which implies that $S_j$ and $\chi_i$ commute for all $i,j$.
All other code checks are either $X$-type or act trivially on $\text{supp}(L)$ and so trivially commute with $\chi_i$ for all $i$.
For all $m$, $\text{supp}(\gamma_m)\in \text{ker}\left(H_{\text{X, gadget}}\right)$ which implies that $\gamma_m$ commutes with $\chi_i$ for all $i$ and $m$.
Finally, for all $m$, $\gamma_m$ acts trivially on the code block and so trivially commute with all code checks.

Additional gadget qubits may be required to increase the expansion of the Tanner graph, which ensures that the code’s distance is preserved during the deformation~\cite{Williamson2024, Ide2025}.
Following Ref.~\cite{Cross2024} we define the boundary Cheeger constant as follows.

\begin{definition} (Boundary Cheeger constant, Ref.~\cite{Cross2024})
For a bipartite graph $F: V \to C$ and a set of vertices $v \subset V$, define the boundary $\partial v \subset C$ to be the set of vertices with an odd number of neighbors in $v$. The boundary Cheeger constant of a Tanner graph is defined as
\begin{equation}
    h = \min_{v \subset V, |v| \leq |V|/2} |\partial v|/|v|
\end{equation}
\end{definition}

Then Theorem~6 in Ref.~\cite{Cross2024} states that whenever the $X$ Tanner graph of the gadget has $h \geq 1$ then the gadget is distance preserving. When every qubit in the gadget is connected to exactly two $X$-type checks, as will be the case for the gadgets constructed in this paper, this reduces to the standard Cheeger constant on the graph whose vertices are $X$-checks and edges are qubits. It is known that the Cheeger constant of a graph on $n$ vertices can be increased by adding $O(n)$ new edges. We will utilise this to ensure our gadgets are distance preserving.

\subsection{Measurement of Arbitrary Logical Operators}
\label{sec:2.2}

A stabiliser code automorphism is a permutation of the physical qubits of the code that commutes with the stabiliser group.
Given the method presented in Sec.~\ref{sec:2.1}, we can now use code automorphisms to present a method for measuring arbitrary logical operators. 

\subsubsection{Measuring Many Logical Operators With One Gadget}
We begin by showing how a single gadget can be used to measure many different logical operators.
This is supported by the following definition.
Note that we use a bar to denote a logical operator's action on the logical space, while an unbarred logical operator denotes the action on the physical qubits.

\begin{definition}
Let $L$ be a logical operator of a stabiliser code and $G$ be a group of code automorphisms of that code. Then, the \textbf{logical orbit of $L$ under G}, $\mathbf{\mathcal{O}_G(L)}$, is defined to be the set of logical actions of elements of the orbit of $L$ under the action of $G$:
\begin{equation}
\mathcal{O}_G(L):=\left\{\overline{g(L)} \mid g\in G\right\}
\end{equation}
\end{definition}

In this context, we call the operator $L$ a \textit{seed operator} as it is used as the seed from which a set of logical operators is generated by the action of code automorphisms. The importance of seed operators is that a gadget constructed for a seed operator can be used to measure any operator in its logical orbit.
Indeed, by definition, the code's Tanner graph is invariant under the action of any code automorphism, $g$.
Therefore, if $\mathcal{M}_L$ is a gadget that measures $L$ when it is connected to code qubits $i\in \text{supp}(L)$ and code checks $S_j\in \mathcal{S}_L$ then the same gadget measures $g(L)$ if it is instead connected to code qubits $g(i)\in\text{supp}(g(L))$ and code checks $g(S_j)\in\mathcal{S}_{g(L)}$.
From this it follows that a single gadget, $\mathcal{M}_L$ can measure any operator in $\mathcal{O}_G(L)$, depending on where it is connected to the code block.

This observation motivates the following definition.

\begin{definition}
A set of seed operators, $\mathcal{L}$, is defined to be a \textbf{complete seed set} (under $G$) if the product of the logical orbits of elements of $\mathcal{L}$ (under $G$) is the full Pauli group on the logical space, i.e.
\begin{equation}
\prod_{L\in\mathcal{L}} \mathcal{O}_G \left(L\right) := \left\{\prod_{L\in\mathcal{L}}\overline{g_L(L)}\middle\vert\{g_L\}_{L\in\mathcal{L}}\subseteq G\right\} =\bar{\mathcal{P}}
\end{equation}
\end{definition}

Given a complete seed set, $\mathcal{L}$ it follows from the above argument that one gadget for each element of $\mathcal{L}$, along with the ability to bridge gadgets to measure a product of logical operators measured by individual gadgets, suffices to measure any logical operator of the code.

\subsubsection{Bridging Gadgets to Measure Arbitrary Logical Operators}
If we have a set of logical operators that can be measured by different gadgets, the product of these operators can be measured by bridging these gadgets~\cite{Cross2024}. 
This involves connecting each gadget to the code block as required to measure each individual logical operator but also to a set of additional bridging qubits.

In the case where $L_1$ and $L_2$ are two logical operators with disjoint support, then they can be bridged to measure $L_1L_2$ using $w$ additional data qubits and $w-1$ additional checks, where $w:=\min(\text{wt}(L_1,L_2))$~\cite{Cross2024}.
Specifically, this is done by connecting each gadget to a common bridge and adding checks to fix the additional gauge operators arising from the addition of the bridge.
The SkipTree algorithm can be used to ensure that this does not significantly increase check weights or qubit degrees~\cite{Swaroop2025}.
The code distance is preserved provided the Cheeger constant of the bridged system satisfies $h\geq 1$; if necessary, extra ancillary qubits and checks can be added to ensure this condition is met~\cite{Williamson2024}.
Iterating this bridging allows for the measurement of $m$ logical operators with disjoint support with the addition of $O(mw)$ qubits, where $w$ is the maximum weight of the logical operators.

In general, logical operators that are measured by different gadgets in our construction can have overlapping support, on a qubit $q$, say.
If the logical operators are not the same Pauli-type, and we want to measure their product, then checks from the different gadgets that have support on $q$ may need to be modified to ensure that they commute with each other.
This can be dealt with by first bridging all the measurement gadgets, and then by merging some of the checks from distinct gadgets that attach to $q$.
In particular, $X$-type checks with support on $q$ should be merged in pairs, leaving either zero or one unmerged $X$-type check with support on $q$, depending on whether there were an even or odd number of $X$-type checks initially.
$Z$-type checks should similarly be merged in pairs.
If there is still an $X$-type check and $Z$-type check left over, they should be merged.
There are two things to check: (i) that all the checks of the resulting gadget commute, and (ii) that the resulting measurement gadget is still distance-preserving.
With regards to (i), consider that all the checks included in the merges already commute with all the other code and gadget checks.
Therefore, the merged checks will still commute with all other code and gadget checks. The merged checks commute with each other since they act either as the identity or as a Pauli-$Y$ operator on their shared support.
With regards to (ii), the fact that the merged measurement gadget is distance-preserving follows since the merge operation does not decrease the Cheeger constant of the graph describing the bridged gadget (assuming we preserve parallel edges in the graph). 

\section{Example Application}
\label{sec:example}
To demonstrate the utility of this construction, we show how it can be applied to a specific family of QLDPC codes to allow for low-overhead quantum computing. 

In particular, we consider a family of generalised bicycle codes~\cite{Kovalev2013,Lin2025}.
Generalised bicycle codes are specified by a lift, $l\in\mathbb{N}$ and two sets $A,B\subseteq\mathbb{Z}_l$.
The physical qubits consists of two sectors, each of size $l$, giving a total of $n=2l$ physical qubits.
Each physical qubit is labelled by $(\rho,\sigma)$ where $\rho\in\mathbb{Z}_l$ denotes the position within the sector and $\sigma\in\{L,R\}$ denotes the sector.

The stabiliser group is defined by the generators:
\begin{align}
S_{X,j} &=\prod_{a\in A} X_{(j+a),L} \prod_{b\in B} X_{(j+b),R}\\
S_{Z,j} &=\prod_{a\in A} Z_{(j-a),R} \prod_{b\in B} Z_{(j-b),L}
\end{align}
for $j\in\mathbb{Z}_l$, where addition and subtraction is mod $l$.
Simultaneous cyclic shifts of the qubits of both sectors by $s$ places map $S_{X,j} \mapsto S_{X,(j+s)}$ and $S_{Z,j} \mapsto S_{Z,(j+s)}$ and thus preserve the stabiliser group.
Therefore, a generalised bicycle code with a lift of $l$ admits a group of automorphisms isomorphic to $\mathbb{Z}_l$ consisting of simultaneous cyclic shifts of the $l$ qubits of each sector.
This means that each seed operator of a generalised bicycle code can have a logical orbit of up to $l$ elements.

\begin{figure}
  \centering
    \includegraphics[width=\linewidth]{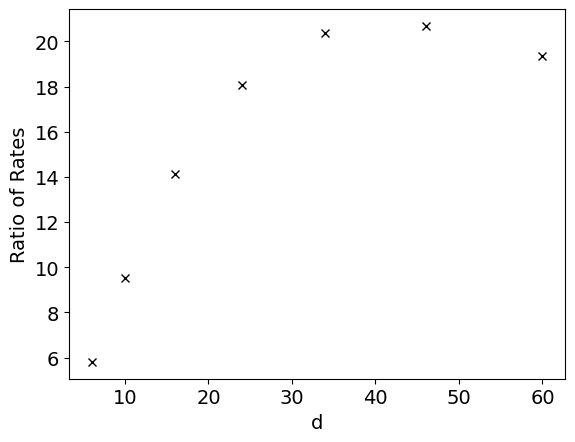}
\caption{Ratio of the rate ($k/n)$ of the $\llbracket 2(2^r-1),2r,r+(r-4)^2 \rrbracket$ generalised bicycle codes to rotated surface codes of the same distance.
  \label{fig:overheads}}
\end{figure}

We consider a family of generalised bicycle codes (parameterised by $r\geq 5$) with $l=2^r-1$, weight six check operators, which we conjecture to have parameters $\llbracket n,k,d \rrbracket=\llbracket 2(2^r-1),2r,r+(r-4)^2 \rrbracket$.
While the rate of such codes ($k/n=r/(2^r-1)$) asymptotically decays exponentially with increasing distance, they perform well for distances relevant to real quantum devices.
Indeed, as shown in Fig.~\ref{fig:overheads}, for $10\leq d\leq 60$, they have a rate that is approximately ten to twenty times greater than that of a rotated surface code of the same distance.

The motivation for focusing on generalised bicycle codes with these parameters is that their $2^k=2^{2r}$ dimensional logical space can be partitioned into two subspaces that each consist of $l=2^r-1$ $\bar{X}$ operators and $\bar{Z}$ operators.
This makes it possible that such codes could admit a complete seed set consisting of just four elements (two logical $X$-type operators and two logical $Z$-type operators).
Indeed, we show by explicit construction that this can be achieved, using the codes and seed operators specified in Appendix.~\ref{app:seed_operators}.
This means that only four small gadgets are required to measure arbitrary logical Pauli operators.

Table.~\ref{tab:overhead} shows the total overhead -- including all data and ancilla qubits in the code block, gadgets and bridges -- for a processing block capable of arbitrary logical Pauli measurements on the first four codes in this family.
It shows that less than $\sim 100$ physical qubits are required per logical qubit for distances up to $d=24$ -- more than an order of magnitude less than for standard surface code architectures.

\begin{table*}[htbp]
\caption{Overhead for processing blocks for each of the four codes in Appendix.~\ref{app:seed_operators}
\textit{Code block qubits} = $2n$, which includes $n$ data qubits and $n$ syndrome ancilla qubits.
\textit{Gadget qubits} denotes the number of qubits (including ancillae) in the gadgets constructed for each seed operator, multiplied by four to account for the four seed operators.
$+n$ indicates the number of qubits that need to be added to ensure the gadgets have boundary Cheeger constant of 1.
\textit{Bridge qubits} denotes the number of qubits (including ancillae) required to construct a bridge for each gadget.
\textit{Total qubits} is the sum of the preceding three columns, and accounts for all data and ancilla qubits required to construct a processing block that can perform arbitrary logical Pauli measurements; $n_{\text{total}}/k$ is this number divided by the number of logical qubits in a code block.
The final column compares $n_{\text{total}}/k$ to the number of physical qubits per logical qubit in a standard surface code architecture (e.g., the baseline architecture of Ref.~\cite{Litinski2023}); this is given by $4d^2$, which accounts for data qubits and syndrome ancilla qubits in two surface code patches per logical qubit -- one storing the logical qubit and the other an ancillary patch required for logical Pauli measurements through lattice surgery.
\label{tab:overhead}}
\resizebox{\textwidth}{!}{
\begin{tabular}{| c | c | c | c | c | c | c |}
\hline
Code & Code Block & Gadget & Bridge & Total & $n_{\text{total}}/{k}$ & Factor Overhead Reduction\\
& Qubits & Qubits & Qubits & Qubits & & vs Surface Code\\
\hline
$\llbracket 62,10,6 \rrbracket$ & 124 & $19\times 4 = 76$ & $11\times 4=44$ & 244 & 24 & 6\\
\hline
$\llbracket 126,12,10 \rrbracket$ & 252 & $31\times 4 = 124$ & $19\times 4=76$ & 452 & 38 & 11\\
\hline
$\llbracket 254,14,16 \rrbracket$ & 508 & $(49+8)\times 4= 228$ & $31\times 4 = 124$ & 860 & 61 & 17\\
\hline
$\llbracket 510,16,24 \rrbracket$ & 1020 & $(79+20) \times 4 = 396$ & $51\times 4 = 204$ & {1620}  & {101} & {23}\\
\hline
\end{tabular}}
\end{table*}

\section{Conclusion}
We have presented an explicit construction for low-overhead gadgets on QLDPC codes.
Application of this construction to a family of generalised bicycle codes shows that it offers the promise of order-of-magnitude reductions in the overhead of quantum computing relative to surface code architectures.
However, our scheme is not specific to this code family.
It is generally applicable to any QLDPC code and will achieve a low-overhead wherever code automorphisms allow for a small complete seed set.
A wide range of QLDPC codes admit code automorphisms, including bivariate bicycle codes~\cite{Bravyi2024} and more general families of two-block group algebra~\cite{Lin2024} and lifted product codes~\cite{Panteleev2022}.
Future work applying our construction to other codes is therefore likely to lead to even larger reductions in overhead with the potential to hasten the arrival of utility-scale quantum computing.

\bibliographystyle{apsrev4-1}
\bibliography{refs}

\clearpage
\onecolumngrid

\appendix

\section{Seed operators} \label{app:seed_operators}

Specific instances of the generalised bicycle code family described in the main text and their seed operators.
For each code, the logical orbit of $\bar{X}_1$ and $\bar{Z}_1$ contains all logical Pauli operators on the first $k/2$ logical qubits, and the logical orbit of $\bar{X}_{\frac{k}{2}+1}$ and $\bar{Z}_{\frac{k}{2}+1}$ contains all logical Pauli operators on the last $k/2$ logical qubits.

\begin{enumerate}[label=\arabic*)]
    \item $l=31$, $A=\{0, {6}, {15}\}$, $B=\{0, {5}, {7}\}$
        \begin{align*}
            &\bar{X}_1 = X(\{1, 6, 8, 10\},L)\,  X(\{11, 26\},R) \\
            &\bar{Z}_1 = Z(\{3, 12, 18, 19\},L)\,  Z(\{11, 18\},R) \\
            &\bar{X}_{6} = X(\{16, 23\},L)\,  X(\{0, 15, 16, 22\},R) \\
            &\bar{Z}_{6} = Z(\{0, 16\},L)\,  Z(\{1, 3, 5, 10\},R) \\
        \end{align*}
    \item $l=63$, $A=\{0, {4}, {37}\}$, $B=\{0, {29}, {49}\}$
        \begin{align*}
            &\bar{X}_1 = X(\{7, 12, 36, 41, 56\},L)\,
            X(\{1, 27, 31, 38, 61\},R) \\
            &\bar{Z}_1 = Z(\{5, 15, 28, 35, 45, 61\},L)\,
            Z(\{1, 11, 54, 57\},R) \\
            &\bar{X}_{7} = X(\{9, 19, 26, 29\},L)\,
            X(\{5, 15, 22, 38, 48, 55\},R) \\
            &\bar{Z}_{7} = Z(\{2, 25, 32, 36, 62\},L)\,
            Z(\{7, 22, 27, 51, 56\},R) \\
        \end{align*}
    \item $l=127$, $A=\{0, {32}, {100}\}$, $B=\{0, {28}, {49}\}$
        \begin{align*}
            &\bar{X}_1 = X(\{28, 47, 55, 75, 103, 114, 124\},L)\,
            X(\{4, 14, 15, 23, 50, 77, 83, 109, 123\},R) \\
            &\bar{Z}_1 = Z(\{1, 24, 33, 51, 60, 65, 107, 119, 124\},L)\,
            Z(\{7, 8, 36, 85, 106, 114, 124\},R) \\
            &\bar{X}_{8} = X(\{3, 31, 32, 42, 52, 60, 81\},L)\,
            X(\{6, 15, 38, 42, 47, 59, 101, 106, 115\},R)  \\
            &\bar{Z}_{8} = Z(\{0, 8, 9, 19, 27, 41, 67, 73, 100\},L)\,
            Z(\{26, 36, 47, 75, 95, 103, 122\},R) \\
        \end{align*}
    \item $l=255$, $A=\{0, 39, 55\}$, $B=\{0, 70, 127\}$
        \begin{align*}
            \bar{X}_1 =\:   & X(\{18, 31, 35, 36, 91, 126, 146, 163, 164, 180, 196, 216, 233, 253\},L) \\
                            & X(\{48, 52, 87, 101, 103, 106, 107, 125, 140, 156, 179, 211\},R) \\
            \bar{Z}_1 =\:   & Z(\{38, 54, 57, 93, 112, 148, 164, 185, 197, 203, 213, 238, 240, 252\},L) \\
                            &Z(\{18, 55, 59, 73, 129, 130, 142, 182, 187, 199, 244, 252\},R) \\
            \bar{X}_{9} =\: & X(\{6, 27, 35, 80, 92, 97, 137, 149, 150, 206, 220, 224\},L) \\
                            &X(\{27, 39, 41, 66, 76, 82, 94, 115, 131, 167, 186, 222, 225, 241\},R) \\
            \bar{Z}_{9} =\: & Z(\{10, 11, 14, 16, 30, 65, 69, 161, 193, 216, 232, 247\},L) \\
                            &Z(\{26, 81, 82, 86, 99, 119, 139, 156, 176, 192, 208, 209, 226, 246\},R) \\
        \end{align*}
\end{enumerate}

\end{document}